\documentclass[a4paper]{jpconf}
\usepackage{graphicx}

\def \lt {\mbox{$<~$} }

\begin{document}
\title{Strangeness in STAR at RHIC}

\author{Shusu Shi (for the STAR collaboration)}

\address{Key Laboratory of Quarks and Lepton Physics (MOE) and Institute of Particle Physics, Central China Normal University, Wuhan, 430079, China}

\ead{shiss@mail.ccnu.edu.cn}

\begin{abstract}
We present the recent results of strangeness production at the mid-rapidity
in Au + Au collisions at RHIC, from $\sqrt{s_{\rm NN}}$ = 7.7 to 200 GeV. 
The $v_2$ of multi-strange baryon $\Omega$ and $\phi$ mesons
are similar to that of pions and protons in the intermediate $p_T$ range (2 - 5 GeV/$c$) in $\sqrt{s_{\rm NN}}$ = 200 GeV Au + Au collisions, 
indicating that the major part of collective flow has been built up at partonic stage. The breaking of mass ordering between $\phi$ mesons
and protons in the low $p_T$ range ($<$ 1 GeV/$c$) is consistent with a picture that $\phi$ mesons are less sensitive to later hadronic interaction.
The nuclear modification factor $R_{\rm CP}$ and baryon to meson ratio change dramatically when the collision energy is
lower than 19.6 GeV. It suggests a possible change of medium property of the system compared to those from high energies. 

\end{abstract}

\section{Introduction}

The main motivations of the experiments of heavy-ion collision at RHIC are 1) generating the new form of matter Quark Gluon Plasma at high energy collisions and studying its properties, and 2) exploring the QCD phase structure by scanning the collision energy from 200 to 7.7 GeV.
Strangeness production is regarded as a sensitive probe to the phase transition, because the strange quark mass is supposed to
be much higher than system temperature in hadronic phase, but lower than system temperature in partonic phase~\cite{introduction}.
In addition, the hadronic interaction cross sections for strange hadrons, especially for multi-strange hadrons $\Xi$, $\Omega$ and $\phi$ mesons are expected to be small and 
their freeze-out temperatures are close to the quark-hadron transition temperature predicted by lattice QCD~\cite{white, multistrange1, multistrange2}. 
Hence, these hadrons are expected to provide information primarily from the partonic stage of the collision. 
They are good probes to study the QGP properties and explore the QCD phase structure.
The STAR experiment has covered the beam energies of $\sqrt{s_{\rm NN}}$  = 7.7, 11.5, 14.5, 19.6, 27, 39, 62.4 and 200 GeV. 
The extracted baryonic chemical potental ($\mu_{B}$) range based on a statistical model~\cite{statmodel} from the 0-5\% central collisions is 
$20\leq\mu_{B}\leq420~{\rm MeV}$ which covers a wide region of the QCD phase diagram.
In these proceedings, we will focus on the following observables for strange and multi-strange hadrons:
elliptic flow $v_2$, nuclear modification factor $R_{\rm CP}$ and baryon to meson ratio.

\section{Results and Discussions}

\begin{figure*}[ht]
\vskip 0cm
\begin{center} \includegraphics[width=0.7\textwidth]{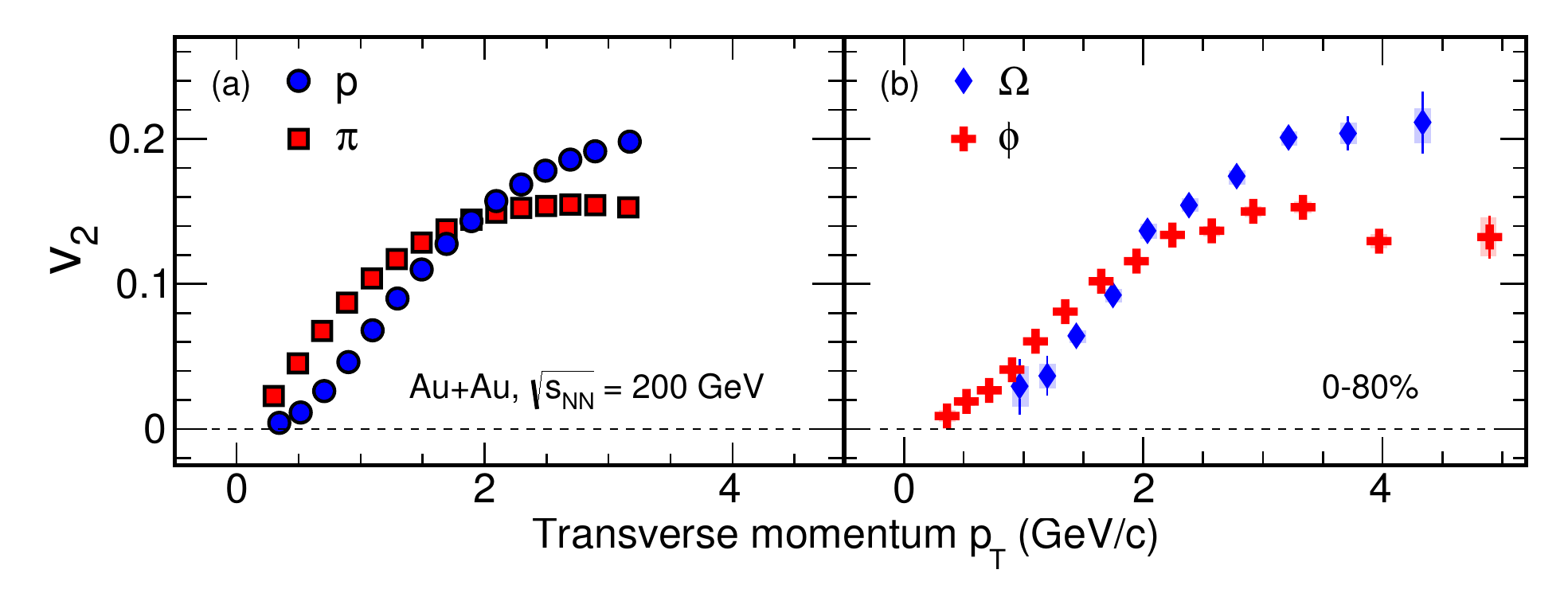}\end{center}
\caption{The $v_{2}$ as function of $p_{T}$ for $\pi$, $p$ (a) and $\phi$, $\Omega$ (b) in Au+Au collisions at $\sqrt{s_{\rm NN}}$ = 200 GeV for 0-80$\%$ centrality~\cite{paper}. }
\label{figure1}
\end{figure*}

\begin{figure*}[ht]
\vskip 0cm
\begin{center}\includegraphics[width=0.6\textwidth]{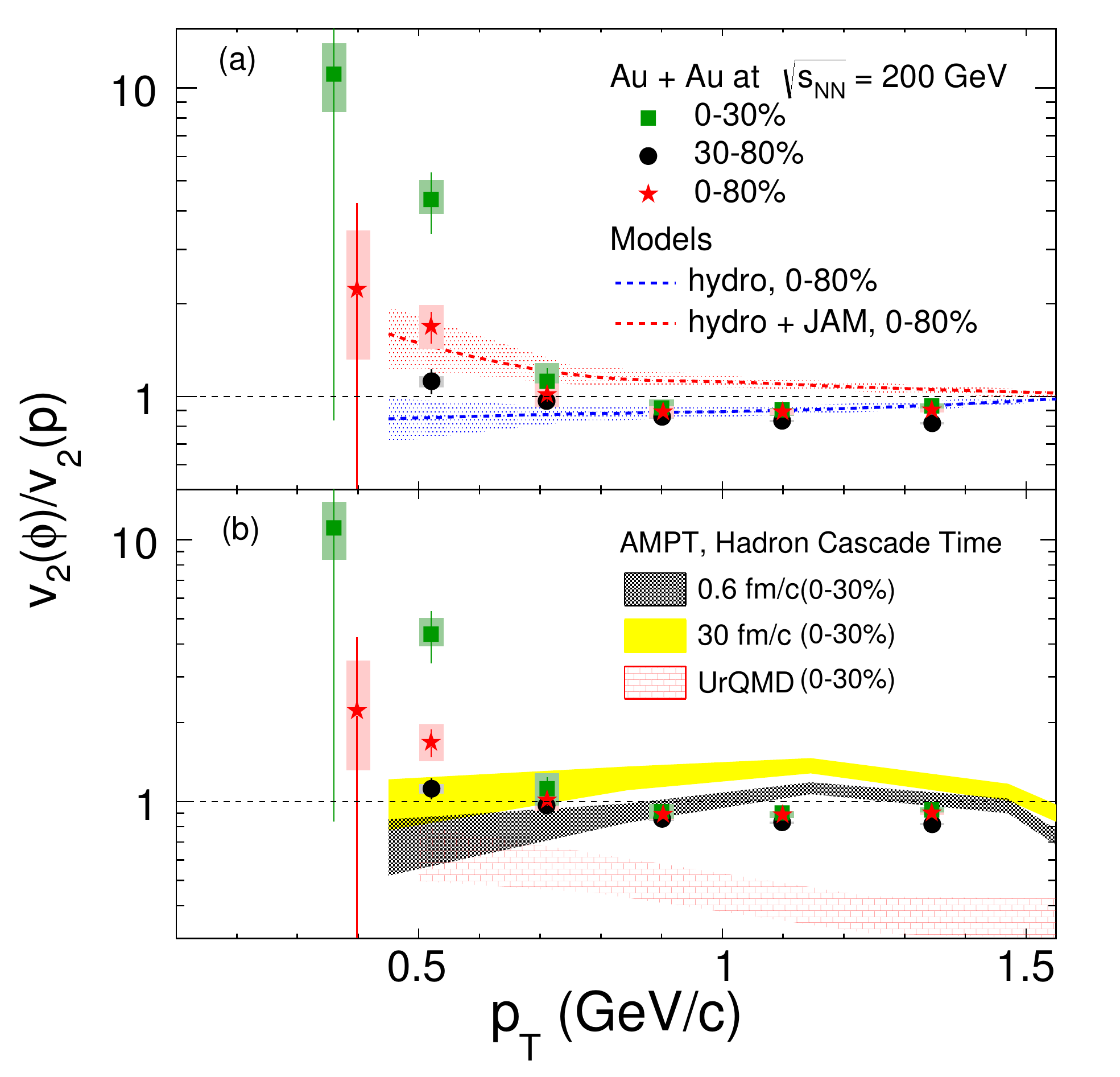}\end{center}
\caption{The ratio of $v_{2}(\phi)$ to $v_{2}(p)$ as function of $p_{T}$ in Au + Au collisions at
$\sqrt{s_{\rm NN}}$ = 200 GeV for  0-30$\%$ and 30-80$\%$ centrality. 
The bands in panel (a) and (b) represent the hydro and transport model calculations for $v_{2}(\phi)/v_{2}(p)$, respectively~\cite{paper}. }
\label{figure2}
\end{figure*}

\begin{figure*}[ht]
\vskip 0cm
\begin{center}\includegraphics[width=0.6\textwidth]{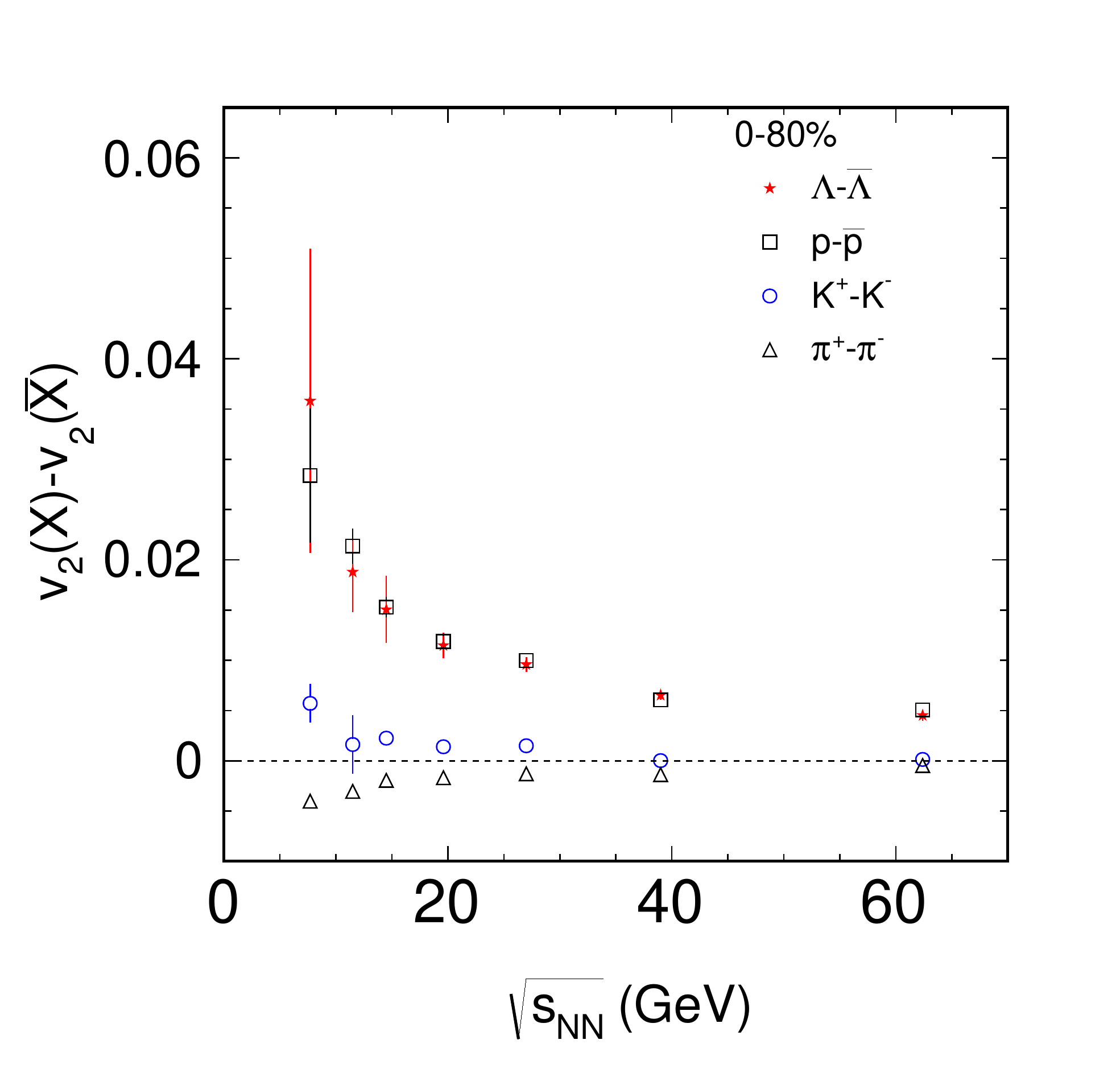}\end{center}
\caption{The difference in $v_2$ between particles
($X$) and their corresponding antiparticles ($\bar{X}$) as a function of beam energy for $0-80$\% central Au + Au collisions~\cite{BES, sss}.}
\label{figure3}
\end{figure*}

In the RHIC runs of year 2010 and 2011, about 730 million minimum bias events of Au + Au collisions  at $\sqrt{s_{\rm NN}}$ = 200 GeV were recorded by STAR.  
Sufficient statistics of multi-strange hadrons and $\phi$ mesons support the precise measurements on $v_2$.
Figure~\ref{figure1} shows the $v_{2}$ as a function of $p_{T}$ for (a) pions, protons  and (b) $\phi$, $\Omega$  in Au + Au collisions at
$\sqrt{s_{\rm NN}}$ = 200 GeV for 0-80\% centrality~\cite{paper}. 
A comparison between $v_{2}$ of pions and protons, consisting of up ($\it u$) and down ($\it d$) light constituent quarks is shown in panel (a). 
Correspondingly, panel (b) shows a comparison of $v_{2}$ of $\phi$ and $\Omega$ containing $\it s$ constituent quarks. 
This is the first time that high precision measurement of $\Omega$ baryon $v_2$ up to 4.5 GeV/$c$ is available in experiments of heavy-ion collisions.
In the low $p_{T}$ region ( $p_{T}$ $<$ 2.0 GeV/$c$), the $v_{2}$ of  $\phi$ and $\Omega$ follows mass ordering.
At intermediate $p_{T}$  ( 2.0 $<$ $p_{T}$ $<$ 5.0 GeV/$c$), a baryon-meson separation is observed.
It is evident that the $v_{2}(p_{T}) $ of hadrons consisting only of strange constituent quarks ($\phi$ and $\Omega$)  is similar to that of light hadrons, pions and protons. 
However the $\phi$ and $\Omega$ do not participate strongly in the hadronic interactions, because of the smaller hadronic cross sections compared to pions and protons. 
It suggests the major part of the collectivity is developed during the partonic phase in high energy heavy-ion collisions.
Experimental measurements indicate partonic collectivity has been built up in top energy heavy-ion collisions at RHIC.
The $\phi$ mesons and protons show different sensitivity to the hadronic rescatterings. 
The $\phi$ mesons are less sensitive to the late hadron hadron interactions than light hadrons due to the smaller hadronic cross section.
Hydrodynamical model calculations predict that $v_{2}$ as a function of $p_{T}$ for different particle species follows mass ordering, 
where the $v_{2}$ of heavier hadrons is lower than that of lighter hadrons. 
Hirano {\it et al.} predict the mass ordering of $v_{2}$ could be broken between $\phi$ mesons and protons at low $p_{ T}$ ($p_{T}$ $\lt$ 1.5 GeV/$c$)
based on a model with ideal hydrodynamics with hadron cascade process~\cite{hyrdo_cascade}.
As the model calculations assign a smaller hadronic cross section for $\phi$ mesons compared to protons, 
the broken mass ordering is regarded as the different hadronic rescattering contributions on the $\phi$ meson and proton $v_2$.
Figure~\ref{figure2} shows the ratios of $\phi$ $v_{2}$ to proton $v_{2}$ from model calculations and experimental data~\cite{paper}. 
This ratio is larger than unity at $p_{T}$ $\sim$ 0.5 GeV/$c$ for 0-30$\%$ centrality. It indicates
breakdown of the expected mass ordering in that momentum range. This could be due to a large effect of hadronic
rescatterings on the proton $v_{2}$. The data of 0-80\% centrality around 0.5 GeV/$c$ quantitatively agrees with hydro + hadron cascade calculations indicated 
by the shaded red band in panel (a) of Fig.~\ref{figure2}, even though there is a deviation in higher $p_T$ bins. 
A centrality dependence of $v_{2}(\phi)$ to $v_{2}(p)$ ratio is observed in the experimental data. The breakdown of mass ordering of $v_{2}$ is more pronounced in 0-30$\%$
central collisions than in 30-80$\%$ peripheral collisions. 
In the central events,  both hadronic and partonic interactions are stronger than in peripheral events.
Therefore,  the larger effect of late stage hadronic interactions relative to the partonic collectivity
produces a greater breakdown of mass ordering in the 0-30$\%$ centrality data than in the 30-80$\%$.
This observation indirectly supports the idea that the $\phi$ meson has a smaller hadronic interaction cross section.  
The ratio of $\phi$ $v_{2}$ to proton $v_{2}$ was also studied by using the transport models AMPT~\cite{ampt} and UrQMD~\cite{urqmd}. 
The panel (b) of Fig.~\ref{figure2} shows the $v_{2}(\phi)$ to $v_{2}(p)$ ratio for 0-30$\%$ centrality from AMPT and UrQMD models.
The black shaded band is from AMPT with a hadronic cascade time of 0.6 fm/$c$ while the yellow
band is for a hadronic cascade time of 30 fm/$c$.  Larger hadronic cascade time is equivalent to stronger hadronic interactions.
It is clear that the $v_{2}(\phi)/v_{2}(p)$ ratio increases with increasing hadronic cascade time. 
This is attributed to a decrease in the proton $v_{2}$ due to an  increase in hadronic rescattering
while the $\phi$ meson $v_{2}$ is less affected. The ratios from the UrQMD model
are much smaller than unity (shown as a brown shaded band in the panel (b) of Fig.~\ref{figure2}).
The UrQMD model lacks partonic collectivity, thus the $\phi$ meson $v_{2}$ is not fully developed.
None of these models could describe the detailed shape of the $p_T$ dependence.

The most striking feature on the $v_2$ measurements from RHIC Beam Energy Scan program is the observation of an energy dependent difference in $v_2$
between particles and their corresponding antiparticles~\cite{{BES}}.
Figure~\ref{figure3} shows the difference in $v_2$ between particles and their corresponding antiparticles as a function of 
beam energy. The difference between baryon and anti-baryon is much more pronounced than difference between mesons.
Proton versus anti-proton and $\Lambda$ versus $\bar{\Lambda}$ show same magnitude of difference.
This difference naturally breaks the number of constituent quark scaling (NCQ) in $v_2$ which is 
regarded as an evidence of partonic collectivity in the top energy heavy-ion collisions at RHIC. It indicates the hadronic degrees of
freedom play a more important role at lower collision energies.

\begin{figure*}[ht]
\vskip 0cm
\begin{center}\includegraphics[width=0.6\textwidth]{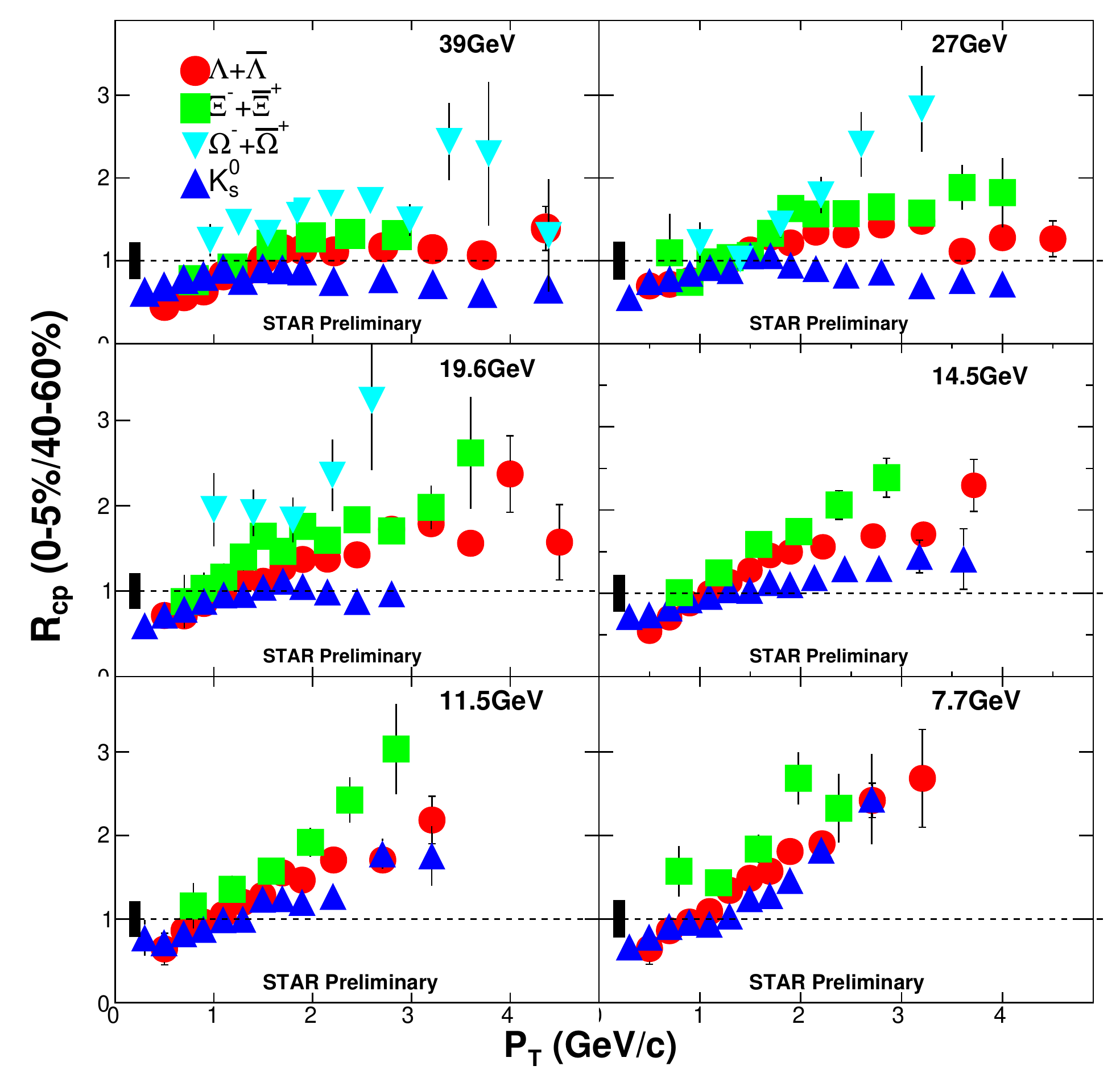}\end{center}
\caption{The nuclear modification factor $R_{\rm CP}$ as a function of $p_T$ in Au + Au collisions at $\sqrt{s_{\rm NN}}$ = 7.7 - 39 GeV
for $K_{S}^{0}$, $\Lambda$, $\Xi$ and $\Omega$ at mid-rapidity $|y| < 0.5$~\cite{usman}.}
\label{figure4}
\end{figure*}

\begin{figure*}[ht]
\vskip 0cm
\begin{center}\includegraphics[width=0.6\textwidth]{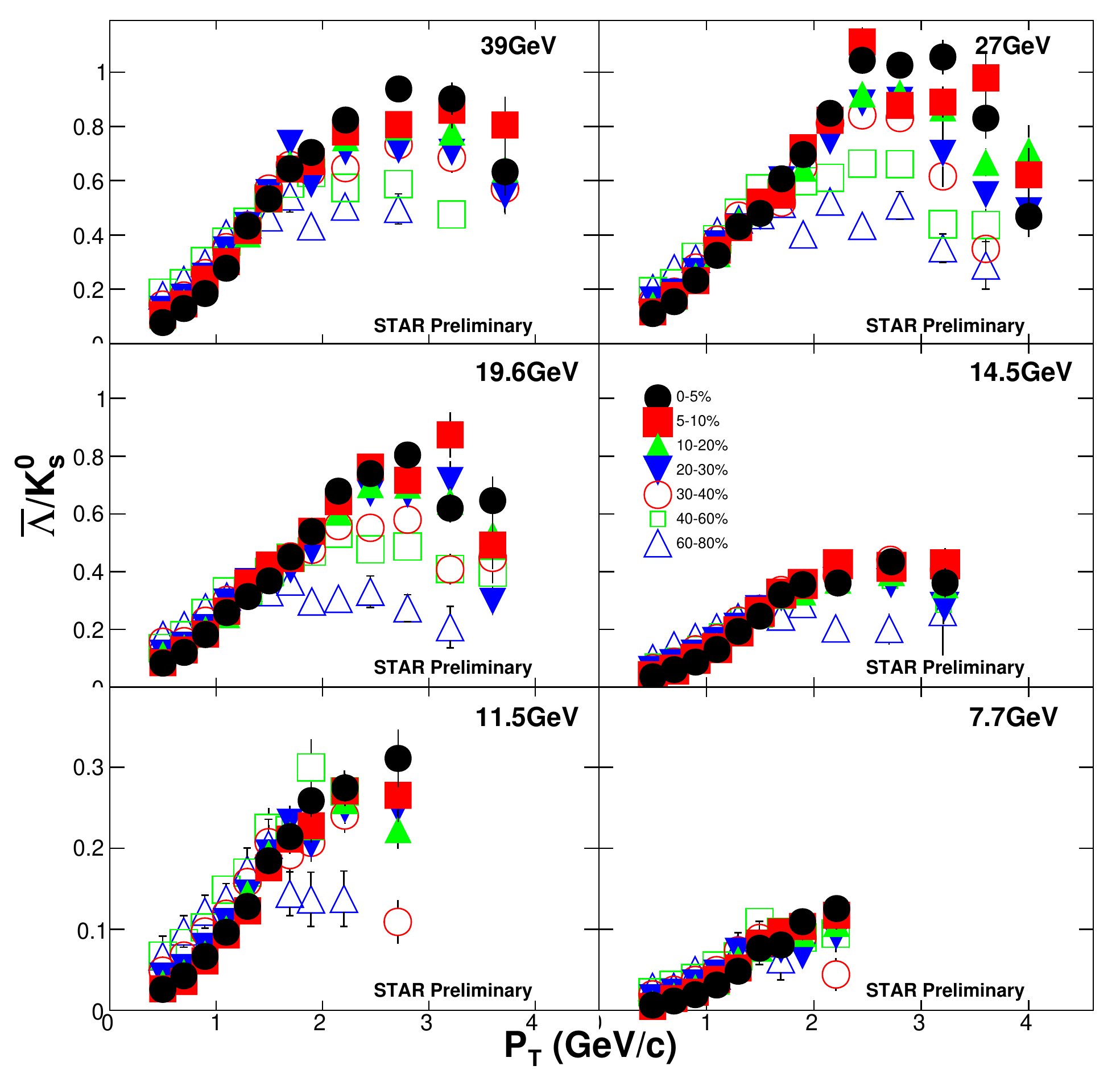}\end{center}
\caption{The ratio of $\bar{\Lambda}/K_{S}^{0}$ as a function of $p_T$ in Au + Au collisions at $\sqrt{s_{\rm NN}}$ = 7.7 - 39 GeV
at mid-rapidity $|y| < 0.5$~\cite{usman}.}
\label{figure5}
\end{figure*}

\begin{figure*}[ht]
\vskip 0cm
\begin{center}\includegraphics[width=0.6\textwidth]{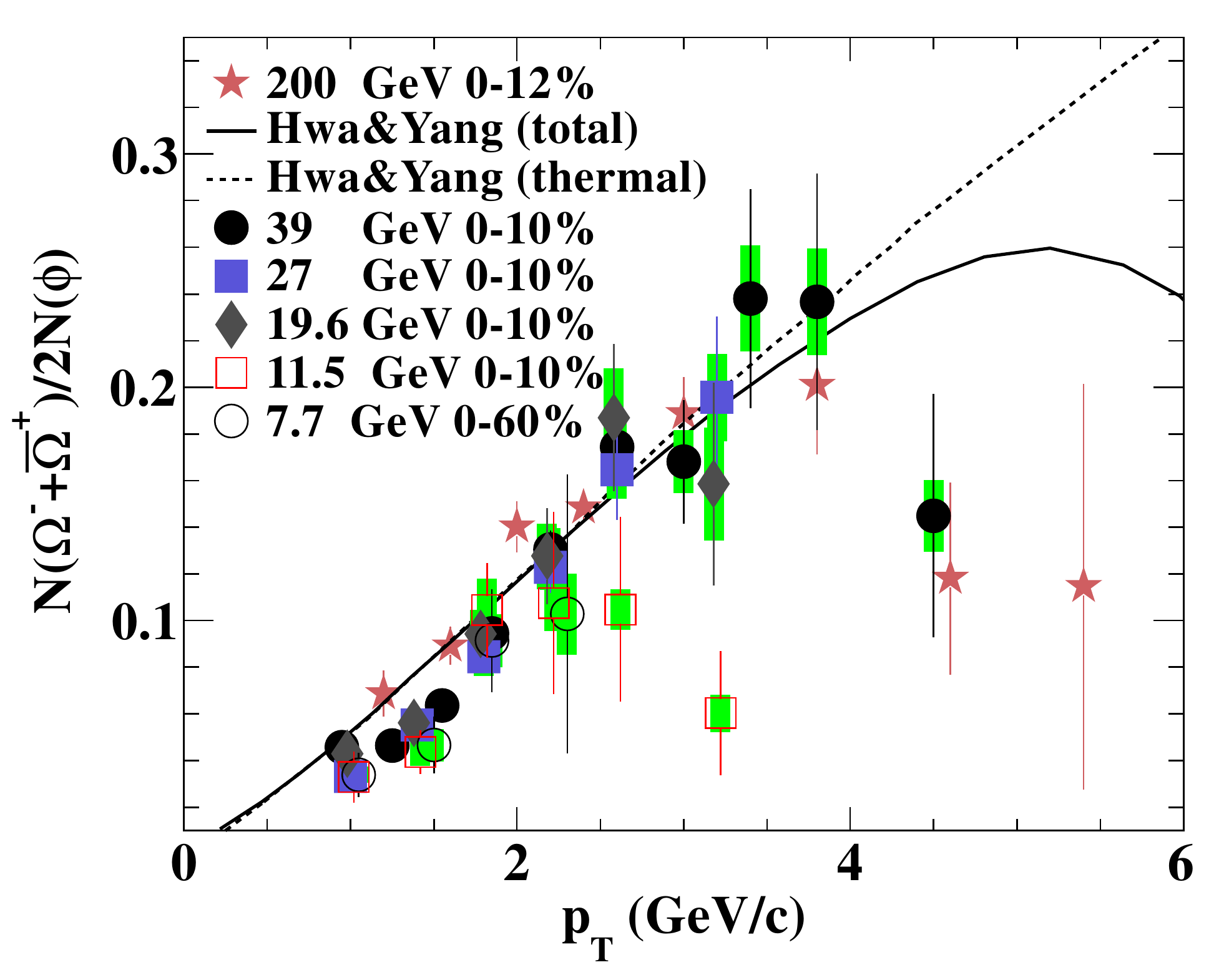}\end{center}
\caption{The ratio of
\emph{N}($\Omega^{-}+\overline{\Omega}^{+}$) to (2\emph{N}($\phi$)) as a function
of $p_{T}$ in  Au+Au collisions at $\sqrt{s_{\rm NN}}$ = 7.7 to 200 GeV at mid-rapidity ($|y| < 0.5$)~\cite{xiaoping, multistrange_pro}.
The solid and dashed lines represent recombination model calculations for central collisions at
$\sqrt{s_{\rm NN}}$ = 200 GeV with
total and thermal strange quark contributions respectively \cite{recombination}.}
\label{figure6}
\end{figure*}

The $R_{\rm CP}$ is defined as the ratios of particle yields in central collisions over those in peripheral ones scaled by the number of inelastic binary  
collisions. Here, $N_{\rm bin}$ is determined from Monte Carlo Glauber model calculations. The $R_{\rm CP}$ of $K_{S}^{0}$, $\Lambda$, $\Xi$ and $\Omega$ in Au+Au 7.7 - 39 GeV collisions are  presented in Fig.~\ref{figure4}~\cite{usman}. 
The $R_{\rm CP}$ will be unity if nucleus-nucleus collisions are just simple superpositions of nucleon-nucleon collisions. Deviation 
of these ratios from unity would imply contributions from nuclear or medium effects. 
At collision energy $\geq$ 19.6 GeV, the $R_{\rm CP}$ of $K_{S}^{0}$
$\leq$ 1 at $p_T$  $>$ 1.5 GeV/$c$ and much less than those of baryons. The high $p_T$ suppression of $K_{S}^{0}$ and
baryon/meson separation is qualitatively consistent with results from RHIC higher energies~\cite{STARBES}. 
However, for collisions at 14.5 GeV and below, the data seem to be qualitatively different from those from higher energies.
There is no suppression for $K_{S}^{0}$ at $p_T$ $>$ 1.5 GeV/$c$, and at intermediate $p_{T}$ the baryon/meson separation becomes less significant. It suggests that 
different properties of the system have been created in collisions at 14.5 GeV and below compared to those from high energies. 

The enhancement of baryon to meson ratios at intermediate $p_T$ compared to elementary collisions is interpreted as a consequence of hadron formation from
parton recombination and parton collectivity~\cite{STARBES}. Therefore, the baryon to meson ratios are expected to be sensitive to parton dynamics of the collision
system. 
Figure~\ref{figure5} shows the ratio of $\bar{\Lambda}/K_{S}^{0}$ as a function of $p_T$ in Au + Au collisions at 7.7 - 39 GeV
at mid-rapidity $|y| < 0.5$~\cite{usman}.
The ratios of $\bar{\Lambda}$ to $K_{S}^{0}$ at intermediate $p_T$ are close to each other at 27 and 39 GeV, and show a slight decrease at 19.6 GeV. There is a 
sudden decrease of intermediate $p_T$ ratios between 19.6 and 14.5 GeV. Besides, the separation of central (0-5\%) and peripheral (40-60\%)
collisions in the ratio becomes less obvious in collisions at 14.5 GeV and below. It suggests a possible change of underlying hadron formation mechanism 
and/or parton dynamics between these two energies.  
We use multi-strange hadrons, $\Omega$ to $\phi$ ratios, to further study the energy dependence of baryon to meson ratios. We present the results as a function
of $p_T$ for Au+Au collisions at 7.7 to 200 GeV in  Fig.~\ref{figure6}~\cite{multistrange_pro}. 
A model calculation by Hwa and Yang for Au+Au collisions at 200 GeV predicted that most of the $\Omega$ and $\phi$ yields up to the intermediate $p_T$ region are from
coalescence of thermal strange quarks~\cite{recombination}. The straight dotted line assumed that these thermal strange quarks have exponential $p_T$
distributions. Deviations from the straight line at high $p_T$ were attributed to recombination with strange quarks from high $p_T$ showers. 
The measured ratios from central Au+Au collisions at 19.6, 27 and 39 GeV follow closely the ratio from 200 GeV and are consistent with a picture 
of coalescence dynamics over a broad $p_T$ range of 1-4 GeV/$c$. The ratios at 11.5 GeV seem to deviate from the trend observed at higher beam 
energies. In particular, the ratios appear to turn down around $p_T$ of 2 GeV/$c$.  The decrease in the $\Omega$ to $\phi$ 
ratios from central collisions at 11.5 GeV compared to those at 19.6 GeV or above may indicate a significant change in the hadron formation and/or in 
strange quark $p_T$ distributions at the lower energy. Such a change may arise from a transition from hadronic to partonic dynamics with increasing beam 
energy.

\section{Summary}
In summary, the high precision $v_2$ data of multi-strange hadrons, especially for $\Omega$ baryon and $\phi$ meson prove the 
partonic collectivity has been built-up at top energy heavy-ion collisions at RHIC. The violation of mass ordering between $\phi$ mesons and protons at low $p_T$ supports 
$\phi$ mesons are less sensitive to late hadronic interactions. $K_{S}^{0}$ $R_{\rm CP}$ increases with decreasing beam energies indicating that partonic energy loss effect is less important at lower energy. The separation of central and peripheral $\bar{\Lambda}/K_{S}^{0}$ ratio is not obvious in collisions at 14.5 GeV and below and the decrease in the $\Omega$ to $\phi$ 
ratios from central collisions at 11.5 GeV suggest the change of medium property and phase transition possibly happen between 19.6 GeV and lower energies.

\section{Acknowledgments}
This work was supported in part by the National Basic Research Program of China (973 program) under grand No. 2015CB8569 and
National Natural Science Foundation of China under grant No. 11475070 and 11628508.


\begin{thebibliography}{00}

\bibitem{introduction} J. Rafelski and B. Muller, Phys. Rev. Lett. \textbf{48}, 1066 (1982).
\bibitem{white} J. Adams {\it et al.}, Nucl. Phys. \textbf{A 757} 102 (2005).
\bibitem{multistrange1} A. Shor, Phys. Rev. Lett. \textbf{54}, 1122 (1985).
\bibitem{multistrange2} H. van Hecke, H. Sorge and N. Xu, Phys. Rev. Lett. \textbf{81}, 5764 (1998).
\bibitem{statmodel} A.~Andronic, {\it et al.}, Nucl. Phys. {\bf A834}, 237(2010); 
A.~Andronic, P.~Braun-Munzinger and J.~Stachel, Nucl. Phys. {\bf A772}, 167(2006).
\bibitem{paper} L. Adamczyk {\it et al.} (STAR Collaboration), Phys. Rev. Lett. \textbf{116}, 062301 (2016).
\bibitem{hyrdo_cascade} T. Hirano  {\it et al.},  Phys. Rev. \textbf{C 77}, 044909 (2008); 
S. Takeuchi {\it et al.}, Phys. Rev. \textbf{C 92}, 044907 (2015).
\bibitem{ampt} Z.-W. Lin {\it et al.},   Phys. Rev. \textbf{C 72}, 064901 (2005).
\bibitem{urqmd} S. A. Bass {\it et al.},  Prog. Part. Nucl. Phys. \textbf{41}, 255 (1998).
\bibitem{BES} L. Adamczyk {\it et al.} (STAR Collaboration), Phys. Rev. C \textbf{86}, 054908 (2012);
Phys. Rev. C \textbf{88}, 014902 (2013);
Phys. Rev. C \textbf{93}, 014907 (2016).
\bibitem{sss} S.S. Shi, arXiv:1607.04863.
\bibitem{usman} M. U. Ashraf (for the STAR Collaboration), J. Phys. Conf. Ser. \textbf{668} 012095 (2016).
\bibitem{STARBES} M. M.Aggarwal {\it et al.} (STAR Collaboration), arXiv:1007.2613.
\bibitem{xiaoping} X. Zhang (for the STAR Collaboration), J. Phys. Conf. Ser. \textbf{668} 012033 (2016).
\bibitem{multistrange_pro} L. Adamczyk {\it et al.} (STAR Collaboration), Phys. Rev. C \textbf{93}, 021903 (2016).
\bibitem{recombination} R. C. Hwa and C. B. Yang, Phys. Rev. C \textbf{66}, 025205 (2002); Phys. Rev. C \textbf{75}, 054904 (2007).

\end{thebibliography}
\end{document}